\documentclass[11pt,twoside,english]{article}
\usepackage[T1]{fontenc}
\usepackage[latin9]{inputenc}
\usepackage[a4paper]{geometry}
\usepackage{color}
\usepackage{float}
\usepackage{mathrsfs}
\usepackage{amsmath}
\usepackage{amssymb}
\usepackage{setspace}
\PassOptionsToPackage{normalem}{ulem}
\usepackage{ulem}
\makeatletter

\DeclareRobustCommand{\greektext}{%
  \fontencoding{LGR}\selectfont\def\encodingdefault{LGR}}
\DeclareRobustCommand{\textgreek}[1]{\leavevmode{\greektext #1}}
\DeclareFontEncoding{LGR}{}{}
\DeclareTextSymbol{\~}{LGR}{126}
\newcommand{\lyxmathsym}[1]{\ifmmode\begingroup\def\b@ld{bold}
  \text{\ifx\math@version\b@ld\bfseries\fi#1}\endgroup\else#1\fi}

\providecommand{\tabularnewline}{\\}

\newcommand{\lyxaddress}[1]{
\par {\raggedright #1
\vspace{1.4em}
\noindent\par}
}

\@ifundefined{date}{}{\date{}}
\makeatother

\usepackage{babel}
\begin{document}

\title{\textbf{Octonionic Gravi-electromagnetism and Dark Matter}}

\author{\textbf{B. C. Chanyal}\textbf{\textit{,}}\textbf{ V. K. Sharma and
O. P. S. Negi}}

\maketitle

\lyxaddress{\begin{center}
\textit{Department of Physics}\\
\textit{Kumaun University, S. S. J. Campus Almora -263601 (U.K.) India}
\par\end{center}}

\lyxaddress{\begin{center}
\textit{Email:- bcchanyal@gmail.com}\\
\textit{~~~~~~~~~~~~~~~~~vishalphysics1985@gmail.com}\\
\textit{~~~~~~~~ops\_negi@yahoo.co.in}
\par\end{center}}
\begin{abstract}
An attempt has been made to analyse the the role of octonions in various
unified field theories associated with dyons and the dark matter.
Starting with the split octonion algebra and its properties, we have
discussed the octonionic unified gauge formulation for $SU(2)\times U(1)$
electroweak theory and $SU(3)\times SU(2)\times U(1)$ grand unified
theory. Describing the octonion eight dimensional space as the combination
of two quaternionic spaces (namely associated with the electromagnetic
interaction (EM-space) and linear gravitational interaction (G-space)),
we have reexamined the unified picture of EM-G space in terms of octonionic
split formulation in consistent manner. Consequently, we have obtained
the various field equations for unified gravi-electromagnetic interactions.
Furthermore, we have reconstructed the field equations of hot and
cold dark matter in terms of split octonions. It is shown that the
difference between the octonion cold dark matter (OCDM) and the octonion
hot dark matter (OHDM) is significant in the formulating of structure
of these two, because the velocities of octonion hot dark matter cause
it to wipe out structure on small scales.

\end{abstract}

\section{Introduction}

According to celebrated Hurwitz theorem there exists \cite{key-1}
a set of four division algebra $\{\mathbb{R},\,\mathbb{C},\,\mathbb{H},\,\mathcal{O}\}$
namely the real numbers ($\mathbb{R}$), complex numbers ($\mathbb{C}$),
quaternions ($\mathbb{H}$) \cite{key-2,key-3} and octonions ($\mathcal{O}$)
\cite{key-4,key-5,key-6}. All these four algebra\textquoteright{}s
are alternative with totally anti symmetric associators. Real number
explains well the classical Newtonian mechanics while the complex
number plays an important role for the explanation beyond the framework
of quantum theory and relativity. Real and complex numbers are limited
only up to two dimensions while quaternions are extended to four dimensions
(one real and three imaginaries), on the other hand the octonions
represent eight dimensions (one real and seven imaginaries). Real
and complex numbers are commutative and associative. Quaternions are
associative but not commutative while the octonions are neither commutative
and nor associative but alternative. Quaternions are well connected
\cite{key-7} with the familiar $SU(2)$ group and naturally unify
\cite{key-8,key-9} electromagnetism and weak force responsible for
the electroweak $SU(2)\times U(1)$ sector of standard model. The
octonion analysis has played an important role in the context of various
physical problems {[}10~-17{]} higher dimensional supersymmetry,
super gravity and super strings. Octonions are also {[}13~-15{]}
used for unification program of strong interaction with successful
gauge theory of fundamental interactions. A theoretical description
of the leptons and quark structure of hadrons has already been proposed
by Günaydin and Gursey \cite{key-18} in the context of octonionic
quantum mechanics in terms of octonionic Hilbert space. 

Rather, there has been continuing interests of many authors {[}19~-21{]}
towards the developments of wave equation and octonion form of Maxwell's
equations. We {[}22~-31{]} have also discussed the significance of
octonions to develop generalized octonion electrodynamics, generalized
split octonion electrodynamics, octonion quantum chromodynamics, octonionic
non-Abelian gauge theory, octonion and conservation laws for dyons,
octonion electrodynamics in isotropic and chiral medium, octonion
massive electrodynamics, octonionic representation of superstring
theory, octonion model of dark matter, octonion symmetric Dirac-Maxwell's
equations and obtained the corresponding field equations and equation
of motion in compact and consistence way. Keeping in mind the recent
update of theoretical physics, we have made to analyze the the role
of octonions in various unified field theories associated with dyons
and the dark matter. Starting with the split octonion algebra and
its properties, we have discussed the octonionic unified gauge formulation
for $SU(2)\times U(1)$ electroweak theory and $SU(3)\times SU(2)\times U(1)$
grand unified theory. Describing the octonion eight dimensional space
as the combination of two quaternionic spaces (namely associated with
the electromagnetic interaction (EM-space) and linear gravitational
interaction (G-space)), we have reexamined the unified picture of
EM-G space in terms of octonionic split formulation in consistent
manner. Consequently, we have obtained the various field equations
for unified gravi-electromagnetic interactions. Furthermore, we have
reconstructed the field equations of hot and cold dark matter in terms
of split octonions. It is shown that the difference between the octonion
cold dark matter (OCDM) and the octonion hot dark matter (OHDM) is
significant in the formulating of structure of these two, because
the velocities of octonion hot dark matter cause it to wipe out structure
on small scales. As such, there exists more cold dark matter than
that of hot dark matter in nature supported by octonionic field equations.

\section{The Split Octonion}

An octonion $\mathcal{O}$ is expressed {[}22-29{]} as a real linear
combination of the unit octonions $(e_{0},e_{1},e_{2},e_{3},$ $e_{4},e_{5},e_{6},e_{7})$
as
\begin{align}
\mathcal{O}\,\,\simeq\,\,(\mathcal{O}_{0},\,\mathcal{O}_{1},\,\mathcal{O}_{2},\,\mathcal{O}_{3},\,\mathcal{O}_{4},\,\mathcal{O}_{5},\,\mathcal{O}_{6},\,\mathcal{O}_{7})\,= & \,\,\,\mathcal{O}_{0}e_{0}+\sum_{A=1}^{7}\mathcal{O}_{A}e_{A}\,,\label{eq:1}
\end{align}
where $e_{A}(A=1,2,.....,7)$ are imaginary octonion units and $e_{0}$
is the real octonion unit element. The octet $(e_{0},e_{1},e_{2},e_{3},e_{4},e_{5},e_{6},e_{7})$
is known as the octonion basis where $e_{A}$ satisfy the following
multiplication rules
\begin{align}
e_{0}= & \,\,1,\,\,\, e_{A}^{2}=-1,\,\,\, e_{0}e_{A}=\,\, e_{A}e_{0}=e_{A},\nonumber \\
e_{A}e_{B}= & -\delta_{AB}e_{0}+f_{ABC}\, e_{C}\,.\,\,\,\,\,(A,B,C=1,2,......7)\label{eq:2}
\end{align}
The structure constants $f_{ABC}$ are completely antisymmetric and
take the value $1$, i.e. 
\begin{equation}
f_{ABC}=+1\longmapsto\,(ABC)\,\equiv\,(123),\,(471),\,(257),\,(165),\,(624),\,(543),\,(736)\,.\label{eq:3}
\end{equation}
Therefore, we get the following relations among octonion basis elements
as
\begin{align}
\left[e_{A},\,\, e_{B}\right]=\,\,2f_{ABC}e_{C}\,,\,\,\, & \left\{ e_{A},\,\, e_{B}\right\} =-\delta_{AB}e_{0}\,,\,\,\,\, e_{A}(e_{B}e_{C})\neq\,\,(e_{A}e_{B})e_{C}\,,\label{eq:4}
\end{align}
where $\delta_{AB}$ is the usual Kronecker delta-Dirac symbol. \\
Like wise, the split octonions \cite{key-25,key-26} are a non associative
extension of split quaternions (associative). Split octonions differ
from the octonions in the signature of quadratic form. The split octonion
have a signature $(4,4)$ whereas the octonions have positive signature
$(8,0)$. The Cayley algebra of octonion over the field of complex
number $\mathbb{C}_{\mathbb{C}}=\mathbb{C}\otimes C$ is visualized
as the algebra of split octonions with its following basis elements,
\begin{equation}
\begin{cases}
u_{0}\,\,=\frac{1}{2}\left(e_{0}+ie_{7}\right)\,; & u_{0}^{*}\,\,=\frac{1}{2}\left(e_{0}-ie_{7}\right)\,;\\
u_{j}\,\,=\frac{1}{2}\left(e_{j}+ie_{j+3}\right)\,; & u_{j}^{*}\,\,=\frac{1}{2}\left(e_{j}-ie_{j+3}\right)\,;\,\,\,\,\,\,\,\,\,(\forall\, j=1,2,3)
\end{cases}\label{eq:5}
\end{equation}
where $i=\sqrt{-1}$ is assumed to commute with $e_{A}(A=1,2,......,7)$
octonion units. The split octonion basis elements $\left(u_{0},\, u_{j},\, u_{0}^{*},\, u_{j}^{*}\right)$
satisfy the following multiplication rules,
\begin{align}
u_{i}u_{j}= & \,\epsilon_{ijk}u_{k}^{*};\,\,\, u_{i}^{*}u_{j}^{*}=-\epsilon_{ijk}u_{k}^{*}\,;\,\,\,\,(\forall i,j,k=1,2,3)\nonumber \\
u_{i}u_{j}^{*}= & -\delta_{ij}u_{0};\,\,\,\,\, u_{i}u_{0}=0;\,\,\,\,\, u_{i}^{*}u_{0}=u_{i}^{*};\nonumber \\
u_{i}^{*}u_{j}= & -\delta_{ij}u_{0};\,\,\,\,\, u_{i}u_{0}^{*}=u_{i};\,\,\,\,\, u_{i}^{*}u_{0}^{*}=0\,;\label{eq:6}\\
u_{0}u_{i}= & \,\, u_{i};\,\,\,\,\,\, u_{0}^{*}u_{i}=0;\,\,\,\,\, u_{0}u_{i}^{*}=0\,;\nonumber \\
u_{0}^{*}u_{i}^{*}= & \,\, u_{i};\,\,\,\,\, u_{0}^{2}=u_{0};\,\,\,\, u_{0}^{*2}=u_{0}^{*};\,\,\, u_{0}u_{0}^{*}=u_{0}^{*}u_{0}=0\,.\nonumber 
\end{align}
Thus the multiplication table of split octonionic basis elements is
given in Table-1.

\begin{table}[H]
\centering{}%
\begin{tabular}{ccccccccc}
\hline 
$\cdot$ & $u_{0}^{*}$ & $u_{1}^{*}$ & $u_{2}^{*}$ & $u_{3}^{*}$ & $u_{0}$ & $u_{1}$ & $u_{2}$ & $u_{3}$\tabularnewline
$u_{0}^{*}$ & $u_{0}^{*}$ & $u_{1}^{*}$ & $u_{2}^{*}$ & $u_{3}^{*}$ & $0$ & $0$ & $0$ & $0$\tabularnewline
$u_{1}^{*}$ & $0$ & $0$ & $u_{3}$ & $-u_{2}$ & $u_{1}^{*}$ & $-u_{0}^{*}$ & $0$ & $0$\tabularnewline
$u_{2}^{*}$ & $0$ & $-u_{3}$ & $0$ & $u_{1}$ & $u_{2}^{*}$ & $0$ & $-u_{0}^{*}$ & $0$\tabularnewline
$u_{3}^{*}$ & $0$ & $u_{2}$ & $-u_{1}$ & $0$ & $u_{3}^{*}$ & $0$ & $0$ & $-u_{0}^{*}$\tabularnewline
$u_{0}$ & $0$ & $0$ & $0$ & $0$ & $u_{0}$ & $u_{1}$ & $u_{2}$ & $u_{3}$\tabularnewline
$u_{1}$ & $u_{1}$ & $-u_{0}$ & $0$ & $0$ & $0$ & $0$ & $u_{3}^{*}$ & $-u_{2}^{*}$\tabularnewline
$u_{2}$ & $u_{2}$ & $0$ & $-u_{0}$ & $0$ & $0$ & $-u_{3}^{*}$ & $0$ & $u_{1}^{*}$\tabularnewline
$u_{3}$ & $u_{3}$ & $0$ & $0$ & $-u_{0}$ & $0$ & $u_{2}^{*}$ & $-u_{1}^{*}$ & $0$\tabularnewline
\hline 
\end{tabular}\caption{Split-Octonion Multiplication Table}
\end{table}

The automorphism group of octonion is described \cite{key-32,key-33}
as $G_{2}$. Its subgroup which leaves imaginary octonion unit $e_{7}$
invariant (or equivalently the idempotents $u_{0}$ and $u_{0}^{*}$)
is $SU(3)$. Here the units $u_{j}$ and $u_{j}^{*}$ ($j=1,2,3$)
transform respectively like a triplet and anti triplet respectively
associated with colour and anti colour triplets of $SU(3)$ group.
We may now introduce a convenient realization for the basis elements
$(u_{0},u_{j},u_{0}^{*},u_{j}^{*})$ in term of Pauli's spin matrices
as
\begin{equation}
\begin{cases}
u_{0}=\left(\begin{array}{cc}
0 & 0\\
0 & 1
\end{array}\right)\,;\qquad & u_{0}^{*}=\left(\begin{array}{cc}
1 & 0\\
0 & 0
\end{array}\right)\,;\\
u_{j}=\left(\begin{array}{cc}
0 & 0\\
e_{j} & 0
\end{array}\right)\,;\qquad & u_{j}^{*}=\left(\begin{array}{cc}
0 & -e_{j}\\
0 & 0
\end{array}\right)\,.\qquad(\forall j=1,2,3)
\end{cases}\label{eq:7}
\end{equation}
So, the split octonion algebra is now expressed in terms of 2$\times$2
Zorn's vector matrices components of which are scalar and vector parts
of a quaternion, i.e.
\begin{align}
\mathcal{O}\,= & \,\,\left\{ \left(\begin{array}{cc}
m & \vec{p}\\
\vec{q} & n
\end{array}\right);\quad m,n\in Sc(\mathbb{H});\quad\&\:\vec{p,}\vec{q}\in Vec(\mathbb{H})\right\} .\label{eq:8}
\end{align}
As such, we write an arbitrary split octonion $A\in\mathcal{O}$ in
terms of following 2$\times$2 Zorn's vector matrix realization as
\begin{align}
A\,=\,\, au_{0}^{*}+bu_{0}+x_{i}u_{i}^{*}+y_{i}u_{i}\,\,\simeq\,\, & \left(\begin{array}{cc}
a & -\overrightarrow{x}\\
\overrightarrow{y} & b
\end{array}\right),\label{eq:9}
\end{align}
where $a$ and $b$ are scalars and $\vec{x}$ and $\vec{y}$ are
three vectors. Thus the product of two octonions in terms of following
2$\times$2 Zorn's vector matrix realization is expressed as
\begin{align}
\left(\begin{array}{cc}
a & \overrightarrow{x}\\
\overrightarrow{y} & b
\end{array}\right)\left(\begin{array}{cc}
c & \overrightarrow{u}\\
\vec{v} & d
\end{array}\right)\,= & \,\left(\begin{array}{cc}
ac+\left(\overrightarrow{x}.\overrightarrow{v}\right) & a\vec{u}+d\vec{x}+\left(\overrightarrow{y}\times\vec{v}\right)\\
c\overrightarrow{y}+b\vec{v}-\left(\vec{x}\times\overrightarrow{u}\right) & bd+\left(\overrightarrow{y}.\overrightarrow{u}\right)
\end{array}\right),\label{eq:10}
\end{align}
where $(\times)$ denotes the usual vector product, $e_{j}\:(j=1,2,3)$
with $e_{j}\times e_{k}=\epsilon_{jkl}e_{l}$ and $e_{j}e_{k}=-\delta_{jk}.$
Octonion conjugate of equation (\ref{eq:9}) in terms of 2$\times$2
Zorn's vector matrix realization is now defined as
\begin{eqnarray}
\overline{A} & \,=\,\, au_{0}+bu_{0}^{\star}-x_{i}u_{i}^{\star}-y_{i}u_{i} & \,\,\simeq\,\,\left(\begin{array}{cc}
b & \overrightarrow{x}\\
-\overrightarrow{y} & a
\end{array}\right).\label{eq:11}
\end{eqnarray}
The norm of $A$ is defined as
\begin{align}
N(A)\,\,=\, & \overline{A}A\,=\, A\overline{A}\,=\,\,(ab+\overrightarrow{x}.\overrightarrow{y})\hat{1}\,=\,\, n(A)\hat{1},\label{eq:12}
\end{align}
where $\hat{1}$ is the identity elements of matrix order $2\times2$,
and the expression $n(A)=(ab+\overrightarrow{x}.\overrightarrow{y})$
defines the quadratic form which admits the composition as
\begin{equation}
n(\vec{A}\cdot\vec{B})\,=\, n(\vec{A})\, n(\vec{B}),\,\,\,\,(\forall\,\vec{A},\vec{B}\in\mathcal{O})\label{eq:13}
\end{equation}
Thus, we may easily express the Euclidean or Minkowski four vector
in split octonion formulation in terms of 2$\times$2 Zorn's vector
matrix realizations. Consequently, any four-vector $A_{\mu}$ (complex
or real) can equivalently be written in terms of the following Zorn
matrix realization as
\begin{eqnarray}
\mathbb{Z}(A) & \,= & \left(\begin{array}{cc}
x_{4} & -\overrightarrow{x}\\
\overrightarrow{y} & y_{4}
\end{array}\right)\,;\,\,\,\,\,\,\,\,\mathbb{Z}(\overline{A})\,=\,\,\left(\begin{array}{cc}
x_{4} & \overrightarrow{x}\\
-\overrightarrow{y} & y_{4}
\end{array}\right).\label{eq:14}
\end{eqnarray}
The split octonion valued space time vector $\mathbb{Z}^{\mu}$ ($\mu=0,1,2,3$)
in terms of the $4\times4$ (space time vector-valued) Zorn matrix
$\mathbb{Z}_{ab}^{\mu}$ may now be expressed as \cite{key-10}, 
\begin{align}
\mathbb{Z}^{\mu}= & x_{0}^{\mu}u_{0}^{*}+y_{0}^{\mu}u_{0}+x_{j}^{\mu}u_{j}^{*}+y_{j}^{\mu}u_{j}\,\,\simeq\,\,\left(\begin{array}{cc}
x_{0}^{\mu}e_{0} & -x_{j}^{\mu}e_{j}\\
y_{j}^{\mu}e_{j} & y_{0}^{\mu}e_{0}
\end{array}\right),\,\,\,\,\,(\forall\, j=1,2,3)\label{eq:15}
\end{align}
where $\mu\,(\forall\,0,1,2,3$) represents the internal four dimensional
space. Here $\mu=0$ describes $U(1)$ abelian gauge structure while
$\mu=j\,\,\,(\forall\, j=1,2,3)$ is associated with the $SU(2)$
non-Abelian gauge structure \cite{key-26}. Here $x_{0}^{\mu},\, x_{j}^{\mu},\, y_{0}^{\mu},\, y_{j}^{\mu}$
are real valued variables for abelian and non-Abelian gauge fields.
When the space time metric \cite{key-10,key-11} is $\eta_{\mu\nu}\hat{1}_{4\times4}$,
the bi linear term
\begin{align}
\frac{1}{4}Trace[\eta_{\mu\nu}\mathbb{Z}^{\mu}\cdot\mathbb{Z}^{\nu}]=\,\, & \frac{1}{4}\eta_{\mu\nu}[x_{0}^{\mu}x_{0}^{\nu}+y_{0}^{\mu}y_{0}^{\nu}+x_{j}^{\mu}x_{j}^{\nu}+y_{j}^{\mu}y_{j}^{\nu}]\, Trace[\hat{1}_{2\times2}]\nonumber \\
=\,\, & \frac{1}{2}\eta_{\mu\nu}[x_{0}^{\mu}x_{0}^{\nu}+y_{0}^{\mu}y_{0}^{\nu}+x_{j}^{\mu}x_{j}^{\nu}+y_{j}^{\mu}y_{j}^{\nu}]\label{eq:16}
\end{align}
describes the inner product. The octonion conjugation is accordingly
defined as
\begin{align}
\mathbb{\overline{Z}}^{\mu}=\,\, & x_{0}^{\mu}u_{0}+y_{0}^{\mu}u_{0}^{*}-x_{j}^{\mu}u_{j}^{*}-y_{j}^{\mu}u_{j}\,\,\simeq\,\,\left(\begin{array}{cc}
y_{0}^{\mu}e_{0} & x_{j}^{\mu}e_{j}\\
-y_{j}^{\mu}e_{j} & x_{0}^{\mu}e_{0}
\end{array}\right),\label{eq:17}
\end{align}
while the Hermitian conjugation is described \cite{key-10,key-11}
as
\begin{align}
(\mathbb{Z}^{\mu})^{\dagger}=\,\, & (x_{0}^{\mu})^{*}u_{0}+(y_{0}^{\mu})^{*}u_{0}^{*}-(x_{j}^{\mu})^{*}u_{j}^{*}-(y_{j}^{\mu})^{*}u_{j}\,\,\simeq\,\,\left(\begin{array}{cc}
(y_{0}^{\mu})^{*}e_{0} & (x_{j}^{\mu})^{*}e_{j}\\
-(y_{j}^{\mu})^{*}e_{j} & (x_{0}^{\mu})^{*}e_{0}
\end{array}\right).\label{eq:18}
\end{align}
The Zorn matrix product of two split octonions 
\begin{align}
\mathcal{A}\,= & \,\,\left(\begin{array}{cc}
A_{0}e_{0} & -A_{i}e_{i}\\
B_{i}e_{i} & B_{0}e_{0}
\end{array}\right)\,\,\,\,\,\,\,\mbox{and}\,\,\,\,\,\,\mathcal{B}\,=\,\,\left(\begin{array}{cc}
C_{0}e_{0} & -C_{i}e_{i}\\
D_{i}e_{i} & D_{0}e_{0}
\end{array}\right)\label{eq:19}
\end{align}
is defined by
\begin{align}
\mathcal{A}\,\mathcal{B}= & \,\left(\begin{array}{cc}
\left(A_{0}C_{0}+A_{j}D_{j}\right)e_{0} & -\left(A_{0}C_{k}+D_{0}A_{k}+\epsilon_{ijk}B_{i}D_{j}\right)e_{k}\\
\left(C_{0}B_{k}+B_{0}D_{k}+\epsilon_{ijk}A_{i}C_{j}\right)e_{k} & \left(B_{0}D_{0}+B_{i}C_{i}\right)e_{0}
\end{array}\right).\label{eq:20}
\end{align}
Accordingly, the octonion $A^{\mu}(x)$ with a space-time index \cite{key-34},
may now be written in terms of the split generators as
\begin{align}
A^{\mu}(x)\,=\,\, & a^{\mu}(x)u_{0}^{*}+b^{\mu}(x)u_{0}+X_{i}^{\mu}(x)u_{i}^{*}+Y_{i}^{\mu}(x)u_{i}\,,\,\,\,\,(i=1,2,3)\label{eq:21}
\end{align}
where the coefficients of $A^{\mu}(x)$ transform like vectors under
space-time transformations. As such, the space-time covariant derivative
of $A^{\mu}(x)$ can be written as the following form:
\begin{align}
A_{+;\alpha}^{\mu}\,=\,\, & A_{\,,\alpha}^{\mu}+\Omega_{\rho\alpha}^{\mu}A^{\rho}\,,\nonumber \\
A_{-;\alpha}^{\mu}\,=\,\, & A_{\,,\alpha}^{\mu}+\Omega_{\alpha\rho}^{\mu}A^{\rho}\,,\label{eq:22}
\end{align}
where $\Omega_{\rho\alpha}^{\mu}\rightarrow\Omega_{\rho\alpha}^{\mu}.\mathbf{1}_{2\times2}$
is the affinity of the nonsymmetric theory \cite{key-34} and $\mathbf{1}_{2\times2}\simeq\left(u_{0}^{*}+u_{0}\right)$
is the unit element of the split octonion algebra. Similarly the space-time
curvature is given by $R_{\mu\nu\rho}^{\sigma}.\mathbf{1}_{2\times2}$,
where $R_{\mu\nu\rho}^{\sigma}$ is the curvature of nonsymmetric
theory.

\section{Octonionic electroweak formulation}

In order to discuss the role of split octonion in particle physics,
we describe the $U(1)\times SU(2)$ electroweak gauge formulation
in terms of the split octonion representation. Thus, we may write
an octonion as the combination of two gauge fields (i.e. $\mathfrak{A}_{\mu}\,\mbox{and\,}\,\mathfrak{B}_{\mu}$
are respectively associated with electric and magnetic charges) expanded
in terms of quaternions, i.e.
\begin{align}
\begin{pmatrix}\mathfrak{A}_{\mu}\\
\mathfrak{B}_{\mu}
\end{pmatrix}\,\,\longmapsto\, & \begin{pmatrix}\mathfrak{A}_{\mu}^{0}e_{0}+\mathfrak{A}_{\mu}^{a}e_{a}\\
\mathfrak{B}_{\mu}^{0}e_{0}+\mathfrak{B}_{\mu}^{a}e_{a}
\end{pmatrix}\,,\,\,\,(\forall\, a=1,2,3)\label{eq:23}
\end{align}
where the components $\mathfrak{A}_{\mu}^{0}$ and $\mathfrak{B}_{\mu}^{0}$
are described respectively as the electric and magnetic four potentials
of dyons for $U(1)$ gauge formalism. Rather, $\mathfrak{A}_{\mu}^{a}$
and $\mathfrak{B}_{\mu}^{a}$ represent two non-Abelian $SU(2)$ gauge
field due to the presence of electric and magnetic charges of dyons.
As such, the covariant derivative for the case of double fold degeneracy
of $U(1)\times SU(2)$ gauge theory is defined as
\begin{align}
\mathscr{D}_{\mu}\,=\,\begin{pmatrix}\mathscr{D}_{\mu}^{\mathsf{e}} & 0\\
0 & \mathscr{D}_{\mu}^{\mathsf{g}}
\end{pmatrix}\,\,\simeq & \,\,\left(\begin{array}{cc}
\partial_{\mu}+\left(\mathfrak{A}_{\mu}^{0}e_{0}+\mathfrak{A}_{\mu}^{a}e_{a}\right) & 0\\
0 & \partial_{\mu}+\left(\mathfrak{B}_{\mu}^{0}e_{0}+\mathfrak{B}_{\mu}^{a}e_{a}\right)
\end{array}\right),\label{eq:24}
\end{align}
where $\mathsf{e}$ and $\mathsf{g}$ represent the electric and magnetic
charges of dyons. From equation (\ref{eq:24}) we get,
\begin{align}
\left[\mathscr{D}_{\mu},\,\mathscr{D}_{\nu}\right]\,= & \,\left(\begin{array}{cc}
G_{\mu\nu}^{0}e_{0}+G_{\mu\nu}^{a}e_{a} & 0\\
0 & \mathscr{G}_{\mu\nu}^{0}e_{0}+\mathscr{G}_{\mu\nu}^{a}e_{a}
\end{array}\right)\,;\label{eq:25}
\end{align}
which is $U(1)\times SU(2)$ octonion gauge field strength for dyons
in terms of 2$\times2$ Zorn matrix vector realization. Equation (\ref{eq:25})
may be reduced to following compact form as 
\begin{align}
\left[\mathscr{D}_{\mu},\,\mathscr{D}_{\nu}\right]\,= & \,\,\mathbb{G}_{\mu\nu},\label{eq:26}
\end{align}
where
\begin{align}
\mathbb{G}_{\mu\nu}= & \,\,\begin{pmatrix}G_{\mu}^{\mathsf{e}} & 0\\
0 & \mathscr{G}_{\mu}^{\mathsf{g}}
\end{pmatrix}\,\,\simeq\,\,\,\left(\begin{array}{cc}
G_{\mu\nu}^{0}e_{0}+G_{\mu\nu}^{a}e_{a} & 0\\
0 & \mathscr{G}_{\mu\nu}^{0}e_{0}+\mathscr{G}_{\mu\nu}^{a}e_{a}
\end{array}\right),\label{eq:27}
\end{align}
with
\begin{equation}
\begin{cases}
G_{\mu\nu}^{0}\,\,= & \partial_{\mu}\mathfrak{A}_{\nu}^{0}-\partial_{\nu}\mathfrak{A}_{\mu}^{0}+e_{0}\left[\mathfrak{A}_{\mu}^{0},\mathfrak{A}_{\nu}^{0}\right]\,,\\
G_{\mu\nu}^{a}\,\,= & \partial_{\mu}\mathfrak{A}_{\nu}^{a}-\partial_{\nu}\mathfrak{A}_{\mu}^{a}+e_{a}\left[\mathfrak{A}_{\mu}^{a},\mathfrak{A}_{\nu}^{a}\right]\,.
\end{cases}\label{eq:28}
\end{equation}
Here $G_{\mu\nu}^{0}$ and $G_{\mu\nu}^{a}$ respectively describe
the abelian and non-Abelian, i.e. $U(1)_{e}\times SU(2)_{e}$ electroweak
gauge structures in presence of electric charge, while
\begin{equation}
\begin{cases}
\mathscr{G}_{\mu\nu}^{0}\,\,= & \partial_{\mu}\mathfrak{B}_{\nu}^{0}-\partial_{\nu}\mathfrak{B}_{\mu}^{0}+e_{0}\left[\mathfrak{B}_{\mu}^{0},\mathfrak{B}_{\nu}^{0}\right]\,,\\
\mathscr{G}_{\mu\nu}^{a}\,\,= & \partial_{\mu}\mathfrak{B}_{\nu}^{a}-\partial_{\nu}\mathfrak{B}_{\mu}^{a}+e_{a}\left[\mathfrak{B}_{\mu}^{a},\mathfrak{B}_{\nu}^{a}\right]\,,
\end{cases}\label{eq:29}
\end{equation}
describes the $U(1)_{m}\times SU(2)_{m}$ electroweak gauge structure
due to the presence of magnetic monopole. From equations (\ref{eq:28})
and (\ref{eq:29}), we get the following gauge field strength of dyons
as
\begin{align}
\begin{pmatrix}G_{\mu\nu}^{0}\\
G_{\mu\nu}^{a}\\
\mathscr{G}_{\mu\nu}^{0}\\
\mathscr{G}_{\mu\nu}^{a}
\end{pmatrix}\,\,\,\Longrightarrow & \,\,\begin{pmatrix}\mathcal{E}_{\mu\nu}^{0}\\
\mathcal{E}_{\mu\nu}^{a}\\
\mathcal{H}_{\mu\nu}^{0}\\
\mathcal{H}_{\mu\nu}^{a}
\end{pmatrix}\,\,.\label{eq:30}
\end{align}
Here $\mathcal{E}_{\mu\nu}^{0}$ and $\mathcal{H}_{\mu\nu}^{0}$ are
defined as the electric and magnetic field tensors of dyons for the
case of $U(1)$ while $\mathcal{E}_{\mu}^{a}$ and $\mathcal{H}_{\mu}^{a}$
describe the $SU(2)$ non-Abelian gauge field strength. Operating
$\mathscr{D}_{\mu}$ given by the equation (\ref{eq:24}) to the $U(1)\times SU(2)$
octonion gauge field strength $\mathbb{G_{\mu\nu}}$ (\ref{eq:27}),
we get
\begin{align}
\mathscr{D}_{\mu}\mathbb{G}_{\mu\nu}\,= & \,\left(\begin{array}{cc}
\partial_{\mu}G_{\mu\nu}^{0}e_{0}+\partial_{\mu}G_{\mu\nu}^{a}e_{a} & 0\\
0 & \mathsf{\partial_{\mu}}\mathscr{G}_{\mu\nu}^{0}e_{0}+\mathsf{\partial_{\mu}}\mathscr{G}_{\mu\nu}^{a}e_{a}
\end{array}\right),\label{eq:31}
\end{align}
which may further be reduced in terms of compect notation of split
octonion formulation, i.e.
\begin{align}
\mathscr{D}_{\mu}\mathbb{G}_{\mu\nu}\,=\,\, & \mathbb{J_{\nu}}\,\,\Longrightarrow\,\,\begin{pmatrix}\mathbb{J}_{\nu}^{\mathsf{e}} & 0\\
0 & \mathbb{J}_{\nu}^{\mathsf{g}}
\end{pmatrix}\,\,\simeq\,\,\left(\begin{array}{cc}
\mathsf{j}_{\nu}^{0}e_{0}+\mathsf{j}_{\nu}^{a}e_{a} & 0\\
0 & \mathsf{k}_{\nu}^{0}e_{0}+\mathsf{k}_{\nu}^{a}e_{a}
\end{array}\right).\label{eq:32}
\end{align}
Here $\mathbb{J_{\nu}}$ is $U(1)\times SU(2)$ form of octonion gauge
current familiar to normal weak current for dyons. From equation (\ref{eq:32}),
we may write the following field equations
\begin{equation}
\begin{cases}
\mathsf{j}_{\nu}^{0}\,\,= & \partial_{\mu}G_{\mu\nu}^{0}\,;\\
\mathsf{j}_{\nu}^{a}\,\,= & \partial_{\mu}G_{\mu\nu}^{a}\,;\\
\mathsf{k}_{\nu}^{0}\,\,= & \partial_{\mu}\mathscr{G}_{\mu\nu}^{0}\,;\\
\mathsf{k}_{\nu}^{a}\,\,= & \partial_{\mu}\mathscr{G}_{\mu\nu}^{a}\,;
\end{cases}\label{eq:33}
\end{equation}
where $\mathsf{j}_{\nu}^{0}$ and $\mathsf{j}_{\nu}^{a}$ are generalized
octonion electroweak current for $U(1)_{e}\times SU(2)_{e}$ (electric
case) while $\mathsf{k}_{\nu}^{0}$ and $\mathsf{k}_{\nu}^{a}$ for
$U(1)_{m}\times SU(2)_{m}$ (magnetic case). Using equation (\ref{eq:30}),
we may obtain the following gauge field equations of dyons for the
case of electroweak gauge formulation,
\begin{equation}
\begin{cases}
\mathsf{j}_{\nu}^{0}\,\,= & \partial_{\mu}\mathcal{E}_{\mu\nu}^{0}\,;\\
\mathsf{j}_{\nu}^{a}\,\,= & \partial_{\mu}\mathcal{E}_{\mu\nu}^{a}\,;\\
\mathsf{k}_{\nu}^{0}\,\,= & \partial_{\mu}\mathcal{H}_{\mu\nu}^{0}\,;\\
\mathsf{k}_{\nu}^{a}\,\,= & \partial_{\mu}\mathcal{H}_{\mu\nu}^{a}\,;
\end{cases}\label{eq:34}
\end{equation}
So, the analogous continuity equation then changes to be
\begin{align}
\mathscr{D}_{\mu}\mathbb{J_{\mu}}\,=\,\left(\begin{array}{cc}
\partial_{\mu}\mathsf{j}_{\mu}^{0}e_{0}+\partial_{\mu}\mathsf{j}_{\mu}^{a}e_{a} & 0\\
0 & \mathsf{\partial_{\mu}}\mathsf{k}_{\nu}^{0}e_{0}+\mathsf{\partial_{\mu}}\mathsf{k}_{\nu}^{a}e_{a}
\end{array}\right)=\,\, & 0.\label{eq:35}
\end{align}
Hence, the split octonion formulation of electroweak gauge theory
is consistent, compect and simpler.

\section{Octonionic formulation of Grand Unified Theory}

Keeping in mind the role of octonion in gauge field theories, let
us start with the local $SU(3)\times SU(2)\times U(1)$ gauge symmetry
which is an internal symmetry representing the Standard Model (SM).
The smallest simple Lie group which contains the standard model, and
upon which the first Grand Unified Theory (GUT) was based, is $SU(5)\supset SU(3)\times SU(2)\times U(1)$.
Thus we may extend $SU(2)\times U(1)$ electroweak gauge theory to
the $SU(3)\times SU(2)\times U(1)$ gauge theory (GUT) in terms of
split octonion formulation. We may now write the $SU(3)\times SU(2)\times U(1)$
gauge field for generalized fields of dyons as
\begin{align}
\begin{pmatrix}\mathfrak{A}_{\mu}\\
\mathfrak{B}_{\mu}
\end{pmatrix}\,\,\,\longmapsto & \,\,\begin{pmatrix}\mathfrak{A}_{\mu}^{0}e_{0}+\mathfrak{A}_{\mu}^{a}e_{a}+\mathfrak{A}_{\mu}^{\alpha}e_{\alpha}\\
\mathfrak{B}_{\mu}^{0}e_{0}+\mathfrak{B}_{\mu}^{a}e_{a}+\mathfrak{B}_{\mu}^{\alpha}e_{\alpha}
\end{pmatrix}\,,\label{eq:36}\\
\,\, & \,(\forall\,\,\mu=0,1,2,3;\,\, a=1,2,3;\,\,\alpha=1,2,......,8)\nonumber 
\end{align}
where the components of electric $\mathfrak{A}_{\mu}^{0}$ and magnetic
$\mathfrak{B}_{\mu}^{0}$ are the four potentials of dyons in case
of $U(1)$. Similarly, the $\mathfrak{A}_{\mu}^{a}$, $\mathfrak{B}_{\mu}^{a}$
and $\mathfrak{A}_{\mu}^{\alpha}$, $\mathfrak{B}_{\mu}^{\alpha}$
describe the $SU(2)$ and $SU(3)$ gauge field strengths for dyons
in terms of split octonions. So, the covariant derivative for $SU(3)\times SU(2)\times U(1)$
octonion gauge field in terms of 2$\times$2 Zorn's vector matrix
realization of split octonions is then described as
\begin{align}
\mathscr{D}_{\mu}\,=\, & \left(\begin{array}{cc}
\partial_{\mu}+\left(\mathfrak{A}_{\mu}^{0}e_{0}+\mathfrak{A}_{\mu}^{a}e_{a}+\mathfrak{A}_{\mu}^{\alpha}e_{\alpha}\right) & 0\\
0 & \partial_{\mu}+\left(\mathfrak{B}_{\mu}^{0}e_{0}+\mathfrak{B}_{\mu}^{a}e_{a}+\mathfrak{B}_{\mu}^{\alpha}e_{\alpha}\right)
\end{array}\right),\label{eq:37}
\end{align}
which is further reduced to
\begin{align}
\left[\mathscr{D}_{\mu},\,\mathscr{D}_{\nu}\right]\,=\, & \left(\begin{array}{cc}
G_{\mu\nu}^{0}e_{0}+G_{\mu\nu}^{a}e_{a}+G_{\mu\nu}^{\alpha}e_{\alpha} & 0\\
0 & \mathscr{G}_{\mu\nu}^{0}e_{0}+\mathscr{G}_{\mu\nu}^{a}e_{a}+\mathscr{G}_{\mu\nu}^{\alpha}e_{\alpha}
\end{array}\right);\label{eq:38}
\end{align}
where $G_{\mu\nu}^{0},\, G_{\mu\nu}^{a},\, G_{\mu\nu}^{\alpha}$ are
defined as the $SU(3)\times SU(2)\times U(1)$ grand-unified gauge
field strengths of dyons in presence of electric charge, whereas $\mathscr{G}_{\mu\nu}^{0},\,\mathscr{G}_{\mu\nu}^{a},\,\mathscr{G}_{\mu\nu}^{\alpha}$
describe the $SU(3)\times SU(2)\times U(1)$ grand-unified gauge field
strengths in presence of magnetic monopole. Hence, equation (\ref{eq:38})
take the following forms
\begin{align}
\left[\mathscr{D}_{\mu},\,\mathscr{D}_{\nu}\right]\,\,= & \,\,\mathbb{G_{\mu\nu}^{\alpha}}\,,\,\,\,(\forall\,\alpha=1,2,......,8)\label{eq:39}
\end{align}
which is $SU(3)\times SU(2)\times U(1)$ octonion gauge field strength
for dyons. Thus, from equation (\ref{eq:39}), we identity the octonion
gauge field components of $SU(3)\times SU(2)\times U(1)$ are
\begin{equation}
\begin{cases}
G_{\mu\nu}^{0}\,\,= & \partial_{\mu}\mathfrak{A}_{\nu}^{0}-\partial_{\nu}\mathfrak{A}_{\mu}^{0}+e_{0}\left[\mathfrak{A}_{\mu}^{0},\mathfrak{A}_{\nu}^{0}\right]\,;\\
G_{\mu\nu}^{a}\,\,= & \partial_{\mu}\mathfrak{A}_{\nu}^{a}-\partial_{\nu}\mathfrak{A}_{\mu}^{a}+e_{a}\left[\mathfrak{A}_{\mu}^{a},\mathfrak{A}_{\nu}^{a}\right]\,;\\
G_{\mu\nu}^{\alpha}\,\,= & \partial_{\mu}\mathfrak{A}_{\nu}^{\alpha}-\partial_{\nu}\mathfrak{A}_{\mu}^{\alpha}+e_{\alpha}\left[\mathfrak{A}_{\mu}^{\alpha},\mathfrak{A}_{\nu}^{\alpha}\right]\,;
\end{cases}\label{eq:40}
\end{equation}
for $U(1)_{e}\times SU(2)_{e}\times SU(3)_{e}$ grand unified gauge
structures in presence of electric charge, and
\begin{equation}
\begin{cases}
\mathscr{G}_{\mu\nu}^{0}\,\,= & \partial_{\mu}\mathfrak{B}_{\nu}^{0}-\partial_{\nu}\mathfrak{B}_{\mu}^{0}+e_{0}\left[\mathfrak{B}_{\mu}^{0},\mathfrak{B}_{\nu}^{0}\right]\,;\\
\mathscr{G}_{\mu\nu}^{a}\,\,= & \partial_{\mu}\mathfrak{B}_{\nu}^{a}-\partial_{\nu}\mathfrak{B}_{\mu}^{a}+e_{a}\left[\mathfrak{B}_{\mu}^{a},\mathfrak{B}_{\nu}^{a}\right]\,;\\
\mathscr{G}_{\mu\nu}^{\alpha}\:= & \partial_{\mu}\mathfrak{B}_{\nu}^{\alpha}-\partial_{\nu}\mathfrak{B}_{\mu}^{\alpha}+e_{\alpha}\left[\mathfrak{B}_{\mu}^{\alpha},\mathfrak{B}_{\nu}^{\alpha}\right]\,;
\end{cases}\label{eq:41}
\end{equation}
are the constituents of $U(1)_{m}\times SU(2)_{m}\times SU(3)_{m}$
gauge structure in presence of magnetic monopole. Thus, octonionic
formulation leads the following gauge field strengths of dyons as
\begin{align}
\begin{pmatrix}G_{\mu\nu}^{0}\\
G_{\mu\nu}^{a}\\
G_{\mu\nu}^{\alpha}
\end{pmatrix}\,\,\Rightarrow\,\,\begin{pmatrix}\mathcal{E}_{\mu\nu}^{0}\\
\mathcal{E}_{\mu\nu}^{a}\\
\mathcal{E}_{\mu\nu}^{\alpha}
\end{pmatrix}\,,\,\,\,\,\,\mbox{and}\,\,\,\, & \begin{pmatrix}\mathscr{G}_{\mu\nu}^{0}\\
\mathscr{G}_{\mu\nu}^{a}\\
\mathscr{G}_{\mu\nu}^{\alpha}
\end{pmatrix}\,\,\Rightarrow\,\,\begin{pmatrix}\mathcal{H}_{\mu\nu}^{0}\\
\mathcal{H}_{\mu\nu}^{a}\\
\mathcal{H}_{\mu\nu}^{\alpha}
\end{pmatrix}\,,\label{eq:42}
\end{align}
where $\mathcal{E}_{\mu\nu}^{0},\,\mathcal{E}_{\mu\nu}^{a},\,\mathcal{E}_{\mu\nu}^{\alpha}$
are respectively defined the electric gauge structures for the case
of $U(1),$ $\, SU(2),\, SU(3)$ gauge field theory, while $\mathcal{H}_{\mu\nu}^{0},\,\mathcal{H}_{\mu\nu}^{a},\,\mathcal{H}_{\mu\nu}^{\alpha}$
are described for magnetic monopole gauge structures of dyons. Now
we operate $\mathscr{D}_{\mu}$ given by the equation (\ref{eq:37})
to the octonion gauge field strength $\mathbb{G_{\mu\nu}^{\alpha}}$
(\ref{eq:39}), we get
\begin{align}
\mathscr{D}_{\mu}\mathbb{G_{\mu\nu}^{\alpha}}\,= & \left(\begin{array}{cc}
\partial_{\mu}G_{\mu\nu}^{0}e_{0}+\partial_{\mu}G_{\mu\nu}^{a}e_{a}+\partial_{\mu}G_{\mu\nu}^{\alpha}e_{\alpha} & 0\\
0 & \mathsf{\partial_{\mu}}\mathscr{G}_{\mu\nu}^{0}e_{0}+\mathsf{\partial_{\mu}}\mathscr{G}_{\mu\nu}^{a}e_{a}+\mathsf{\partial_{\mu}}\mathscr{G}_{\mu\nu}^{\alpha}e_{\alpha}
\end{array}\right),\label{eq:43}
\end{align}
which may further be reduced in terms of compect notation of octonion
form as
\begin{align}
\mathscr{D}_{\mu}\mathbb{G}_{\mu\nu}^{\alpha}\,\,= & \,\,\mathbb{J_{\nu}^{\alpha}}.\label{eq:44}
\end{align}
Here $\mathbb{J_{\nu}^{\alpha}}$ is $U(1)\times SU(2)\times SU(3)$
grand-unified form of octonion gauge current for dyons which may be
expressed in terms of 2$\times2$ Zorn vector matrix realization of
split octonions as
\begin{align}
\mathbb{J_{\nu}^{\alpha}}\,\,= & \,\left(\begin{array}{cc}
\mathsf{j}_{\nu}^{0}e_{0}+\mathsf{j}_{\nu}^{a}e_{a}+\mathsf{j}_{\nu}^{\alpha}e_{\alpha} & 0\\
0 & \mathsf{k}_{\nu}^{0}e_{0}+\mathsf{k}_{\nu}^{a}e_{a}+\mathsf{k}_{\nu}^{\alpha}e_{\alpha}
\end{array}\right)\,.\label{eq:45}
\end{align}
The octonionic formulation (\ref{eq:44}) thus leads to the following
field equations of dyons
\begin{equation}
\begin{cases}
\mathsf{j}_{\nu}^{0}\,\,= & \partial_{\mu}\mathcal{E}_{\mu\nu}^{0};\,\,\,\,(\forall\,\mu,\nu=0,1,2,3)\\
\mathsf{j}_{\nu}^{a}\,\,= & \partial_{\mu}\mathcal{E}_{\mu\nu}^{a};\,\,\,\,(\forall\, a=1,2,3)\\
\mathsf{j}_{\nu}^{\alpha}\,\,= & \partial_{\mu}\mathcal{E}_{\mu\nu}^{\alpha};\,\,\,\,(\forall\,\alpha=1,2,3,....,8)\\
\mathsf{k}_{\nu}^{0}\,\,= & \partial_{\mu}\mathcal{H}_{\mu\nu}^{0};\,\,\,\,(\forall\,\mu,\nu=0,1,2,3)\\
\mathsf{k}_{\nu}^{a}\,\,= & \partial_{\mu}\mathcal{H}_{\mu\nu}^{a};\,\,\,\,(\forall\, a=1,2,3)\\
\mathsf{k}_{\nu}^{\alpha}\,\,= & \partial_{\mu}\mathcal{H}_{\mu\nu}^{\alpha}\,;\,\,\,\,(\forall\,\alpha=1,2,3,....,8)
\end{cases}\label{eq:46}
\end{equation}
were $\mathsf{j}_{\nu}^{0}$ is the $U(1)$ current for electric charge,
$\mathsf{j}_{\nu}^{a}$ is the $SU(2)$ week current associated with
electric charge and $\mathsf{j}_{\nu}^{\alpha}$ is the current associated
with $SU(3)_{c}$ used for chromo electric charge (i.e. normal color).
On the other hand, $\mathsf{k}_{\nu}^{0}$ is $U(1)$ the counterpart
of the four current, $\mathsf{k}_{\nu}^{a}$ is the $SU(2)$ weak
current while the $\mathsf{k}_{\nu}^{\alpha}$ is $SU(3)_{c}$ gluonic
current due to the presence of magnetic monopole (chromo-magnetic
charges or magnetic color). As such, the octonionic formulation, regardless
a generalization of GUTs for the mixing of gauge currents used for
$U(1),\, SU(2)$ and $SU(3)_{c}$ sectors associated respectively
with the electromagnetic, weak and strong interactions in presence
of dyons, also shows the duality invariance as well. Consequently,
the continuity equation is generalized as
\begin{align}
\mathscr{D}_{\mu}\mathbb{J_{\mu}^{\alpha}}\,= & \left(\begin{array}{cc}
\partial_{\mu}\mathsf{j}_{\mu}^{0}e_{0}+\partial_{\mu}\mathsf{j}_{\mu}^{a}e_{a}+\partial_{\mu}\mathsf{j}_{\mu}^{\alpha}e_{\alpha} & 0\\
0 & \mathsf{\partial_{\mu}}\mathsf{k}_{\nu}^{0}e_{0}+\mathsf{\partial_{\mu}}\mathsf{k}_{\nu}^{a}e_{a}+\mathsf{\partial_{\mu}}\mathsf{k}_{\nu}^{\alpha}e_{\alpha}
\end{array}\right)=\,\,0.\label{eq:47}
\end{align}
which leading to the conservation of Noetherian current.

\section{Octonion formulation with linear gravity}

Here we assume the octonion eight dimensional space, combination of
two quaternionic spaces respectively associated with the electromagnetic
interaction (EM-space) and linear gravitational interaction (G-space)
\cite{key-35,key-36,key-37}. Thus, we may write the split octonionic
unified representation of gravitational-electromagnetic (G-EM) space
in terms of following 2$\times$2 Zorn's vector matrix realization
as
\begin{align}
\mathcal{O}\,= & \,\left(\mathcal{O}_{em-space}\,,\,\,\mathcal{O}_{g-space}\right)\nonumber \\
= & \,\,\mathcal{\mathcal{O}}_{em}^{0}u_{0}^{*}+\mathcal{\mathcal{O}}_{g}^{0}u_{0}+\mathcal{\mathcal{O}}_{em}^{j}u_{j}^{*}+\mathcal{\mathcal{O}}_{g}^{j}u_{j}\,\,\simeq\,\,\left(\begin{array}{cc}
\mathcal{O}_{em}^{0}e_{0} & -\mathcal{O}_{em}^{j}e_{j}\\
\mathcal{O}_{g}^{j}e_{j} & \mathcal{O}_{g}^{0}e_{0}
\end{array}\right)\,.\,\,\,\,\,(\forall\, j=1,2,3)\label{eq:48}
\end{align}
As such, any physical quantity $\mathcal{X}\in\mathcal{O}$ may be
written as
\begin{align}
\mathcal{X}\,\,=\,\left(\mathcal{X}_{em}\,,\,\mathcal{X}_{g}\right)\,\,=\,\,\mathcal{X}_{em}^{0}u_{0}^{*}+\mathcal{X}_{g}^{0}u_{0}+\mathcal{X}_{em}^{j}u_{j}^{*}+\mathcal{X}_{g}^{j}u_{j}\,\,\simeq\,\, & \left(\begin{array}{cc}
\mathcal{X}_{em}^{0}e_{0} & -\mathcal{X}_{em}^{j}e_{j}\\
\mathcal{X}_{g}^{j}e_{j} & \mathcal{X}_{g}^{0}e_{0}
\end{array}\right),\label{eq:49}
\end{align}
and accordingly the split octonion differential operator $\mathfrak{D}$
may also be written for unified picture of the EM-G space \cite{key-30}
in terms of 2$\times$2 Zorn's matrix realization \cite{key-23,key-25},
i.e.
\begin{align}
\mathfrak{D}\,\,=\,\,\left(\mathfrak{D}_{em}\,,\,\mathfrak{D}_{g}\right)\,\,\simeq & \,\,\left(\begin{array}{cc}
\mathcal{\partial}_{em}^{0}e_{0} & -\mathcal{\partial}_{em}^{j}e_{j}\\
\mathcal{\partial}_{g}^{j}e_{j} & -\mathcal{\partial}_{g}^{0}e_{0}
\end{array}\right),\label{eq:50}
\end{align}
while the octonion conjugate of equation (\ref{eq:50}) is described
as
\begin{align}
\overline{\mathfrak{D}}\,\,=\,\,\left(\mathfrak{\overline{D}}_{em}\,,\,\overline{\mathfrak{D}}_{g}\right)\,\,\simeq & \,\,\left(\begin{array}{cc}
-\mathcal{\partial}_{em}^{0}e_{0} & \mathcal{\partial}_{em}^{j}e_{j}\\
-\mathcal{\partial}_{g}^{j}e_{j} & \mathcal{\partial}_{g}^{0}e_{0}
\end{array}\right).\label{eq:51}
\end{align}
Like wise, the split octonion valued potential $\mathscr{V}$ for
the unified EM-G space is defined as 
\begin{align}
\mathscr{V}\,\,=\,\,\left(\mathscr{V}_{-}\,,\,\,\mathscr{V}_{+}\right)\,\,=\,\, & \mathscr{V}_{-}^{0}u_{0}^{*}+\mathscr{V}_{+}^{0}u_{0}+\mathscr{V}_{-}^{j}u_{j}^{*}+\mathscr{V}_{+}^{j}u_{j}\,\nonumber \\
=\,\, & \left(\begin{array}{cc}
\mathscr{V}_{-}^{0}e_{0} & -\mathscr{V}_{+}^{j}e_{j}\\
\mathscr{V}_{-}^{j}e_{j} & \mathscr{V}_{+}^{0}e_{0}
\end{array}\right)\,,\label{eq:52}
\end{align}
where we have expressed the split octonion valued potential in terms
two four potential form as the combination of linear gravitational
potential ($\mathscr{V}_{g}^{\mu}$) and electromagnetic potential
($\mathscr{V}_{em}^{\mu}$) for unified EM-G space, i.e.
\begin{align}
\mathscr{V}\,\,=\,\left(\mathscr{V}_{g}\,,\,\,\mathscr{V}_{em}\right)\,\,\simeq & \,\,\left(\begin{array}{cc}
\left(\mathscr{V}_{g}^{0}-\mathscr{V}_{em}^{0}\right)e_{0} & -\left(\mathscr{V}_{em}^{j}+\mathscr{V}_{g}^{j}\right)e_{j}\\
\left(\mathscr{V}_{em}^{j}-\mathscr{V}_{g}^{j}\right)e_{j} & \left(\mathscr{V}_{g}^{0}+\mathscr{V}_{em}^{0}\right)e_{0}
\end{array}\right)\,,\label{eq:53}
\end{align}
where $\mathscr{V}_{g}^{\mu}=\left(\mathscr{V}_{g}^{0},\,\mathscr{V}_{g}^{j}\right)$
and $\mathscr{V}_{em}^{\mu}=\left(\mathscr{V}_{em}^{0},\,\mathscr{V}_{em}^{j}\right)\,\,(\mu=0,1,2,3)$
are respectively denoted as linear gravitational (G-space) and electromagnetic
(EM-space) four potentials. Thus, we may write the split octonion
potential wave equation for unified EM-G space by operating $\overline{\mathfrak{D}}$
given by equation (\ref{eq:51}) to octonion potential $\mathscr{V}$
(\ref{eq:53}) in the following manner,
\begin{align}
\overline{\mathfrak{D}}\mathscr{V}\,\,= & \,\left(\begin{array}{cc}
[-(\mathcal{\partial}_{em}^{0}\mathscr{V}_{g}^{0}+\mathcal{\partial}_{em}^{j}\mathscr{V}_{g}^{j})+ & [(\mathcal{\partial}_{em}^{k}\mathscr{V}_{g}^{0}+\mathcal{\partial}_{em}^{0}\mathscr{V}_{g}^{k}+\epsilon_{ijk}\mathcal{\partial}_{g}^{i}\mathscr{V}_{em}^{j})+\\
(\mathcal{\partial}_{em}^{0}\mathscr{V}_{em}^{0}+\mathcal{\partial}_{em}^{j}\mathscr{V}_{em}^{j})]e_{0}\, & (\mathcal{\partial}_{em}^{k}\mathscr{V}_{em}^{0}+\mathcal{\partial}_{em}^{0}\mathscr{V}_{em}^{k}-\epsilon_{ijk}\mathcal{\partial}_{g}^{i}\mathscr{V}_{g}^{j})]e_{k}\\
\, & \,\\
{}[(-\mathcal{\partial}_{g}^{k}\mathscr{V}_{g}^{0}-\mathcal{\partial}_{g}^{0}\mathscr{V}_{g}^{k}+\epsilon_{ijk}\mathcal{\partial}_{em}^{i}\mathscr{V}_{em}^{j})+ & [(\mathcal{\partial}_{g}^{0}\mathscr{V}_{g}^{0}+\mathcal{\partial}_{g}^{j}\mathscr{V}_{g}^{j})+\\
(\mathcal{\partial}_{g}^{k}\mathscr{V}_{em}^{0}+\mathcal{\partial}_{g}^{0}\mathscr{V}_{em}^{k}+\epsilon_{ijk}\mathcal{\partial}_{em}^{i}\mathscr{V}_{g}^{j})]e_{k} & (\mathcal{\partial}_{g}^{0}\mathscr{V}_{em}^{0}+\mathcal{\partial}_{g}^{j}\mathscr{V}_{em}^{j})]e_{0}
\end{array}\right)\,,\label{eq:54}
\end{align}
which can further be reduced to
\begin{align}
\overline{\mathfrak{D}}\mathscr{V}\,= & \,\,\mathscr{F}\,\,\simeq\,\,\mathscr{F}_{-}^{0}u_{0}^{*}+\mathscr{\mathscr{F}}_{+}^{0}u_{0}+\mathscr{\mathscr{F}}_{-}^{j}u_{j}^{*}+\mathscr{\mathscr{F}}_{+}^{j}u_{j}\,\nonumber \\
=\,\, & \left(\begin{array}{cc}
\mathscr{\mathscr{F}}_{-}^{0}e_{0} & -\mathscr{F}_{+}^{j}e_{j}\\
\mathscr{\mathscr{F}}_{-}^{j}e_{j} & \mathscr{F}_{+}^{0}e_{0}
\end{array}\right)\,\simeq\,\left(\begin{array}{cc}
\left(\mathscr{\mathscr{F}}_{g}^{0}-\mathscr{\mathscr{F}}_{em}^{0}\right)e_{0} & -\left(\mathscr{\mathscr{F}}_{em}^{j}+\mathscr{\mathscr{F}}_{g}^{j}\right)e_{j}\\
\left(\mathscr{\mathscr{F}}_{em}^{j}-\mathscr{\mathscr{F}}_{g}^{j}\right)e_{j} & \left(\mathscr{\mathscr{F}}_{g}^{0}+\mathscr{\mathscr{F}}_{em}^{0}\right)e_{0}
\end{array}\right),\label{eq:55}
\end{align}
where $\mathscr{F}$ is an split octonion which reproduces the field
strength of generalized gravitational-electromagnetic interactions
of dyons. Thus the unified components of $\mathscr{F}$ are expressed
as
\begin{align}
\left(\mathcal{\partial}_{em}^{0}\mathscr{V}_{g}^{0}+\mathcal{\partial}_{em}^{j}\mathscr{V}_{g}^{j}\right)e_{0} & =\,\,\mathscr{F}_{(em-g)}^{j}\,;\nonumber \\
\left(\mathcal{\partial}_{em}^{0}\mathscr{V}_{em}^{0}+\mathcal{\partial}_{em}^{j}\mathscr{V}_{em}^{j}\right)e_{0} & =\,\,\mathscr{F}_{(em-em)}^{j}\,;\nonumber \\
(\mathcal{\partial}_{em}^{k}\mathscr{V}_{g}^{0}+\mathcal{\partial}_{em}^{0}\mathscr{V}_{g}^{k}+\epsilon_{ijk}\mathcal{\partial}_{g}^{i}\mathscr{V}_{em}^{j})e_{k} & =\,\,\mathscr{F}_{(em-g)/(g-em)}^{k}\,;\nonumber \\
(\mathcal{\partial}_{em}^{k}\mathscr{V}_{em}^{0}+\mathcal{\partial}_{em}^{0}\mathscr{V}_{em}^{k}-\epsilon_{ijk}\mathcal{\partial}_{g}^{i}\mathscr{V}_{g}^{j})e_{k} & =\,\,\mathscr{F}_{(em-em)/(g-g)}^{k}\,;\nonumber \\
(-\mathcal{\partial}_{g}^{k}\mathscr{V}_{g}^{0}-\mathcal{\partial}_{g}^{0}\mathscr{V}_{g}^{k}+\epsilon_{ijk}\mathcal{\partial}_{em}^{i}\mathscr{V}_{em}^{j})e_{k} & =\,\,\mathscr{F}_{(g-g)/(em-em)}^{k}\,;\label{eq:56}\\
(\mathcal{\partial}_{g}^{k}\mathscr{V}_{em}^{0}+\mathcal{\partial}_{g}^{0}\mathscr{V}_{em}^{k}+\epsilon_{ijk}\mathcal{\partial}_{em}^{i}\mathscr{V}_{g}^{j})e_{k} & =\,\,\mathscr{F}_{(g-em)/(em-g)}^{k}\,;\nonumber \\
\left(\mathcal{\partial}_{g}^{0}\mathscr{V}_{g}^{0}+\mathcal{\partial}_{g}^{j}\mathscr{V}_{g}^{j}\right)e_{0} & =\,\,\mathscr{F}_{(g-g)}^{j}\,;\nonumber \\
\left(\mathcal{\partial}_{g}^{0}\mathscr{V}_{em}^{0}+\mathcal{\partial}_{g}^{j}\mathscr{V}_{em}^{j}\right)e_{0} & =\,\,\mathscr{F}_{(g-em)}^{j}\,.\nonumber 
\end{align}
Here the terms ($\mathscr{F}_{em-em}^{j}\,,\,\mathscr{F}_{g-g}^{j}$)
are defined the interactions of electromagnetic- electromagnetic (EM-EM),
gravitational- gravitational (G-G) field strength, and $\mathscr{F}_{(em-g)}^{j}\,,\,\mathscr{F}_{g-em}^{j}$
are described as the combinations of electromagnetic- gravitational
(EM-G), gravitational- electromagnetic (G-EM) field strength while
the other components $\mathscr{F}_{(em-g)/(g-em)}^{k},\,\mathscr{F}_{(em-em)/(g-g)}^{k},$
$\,\mathscr{F}_{(g-g)/(em-em)}^{k}\,,$ $\mathscr{F}_{(g-em)/(em-g)}^{k}$
are defined the combined interactions of electromagnetic and gravitational
for octonion space. Applied the following Lorentz Gauge conditions
\begin{equation}
\begin{split}\left(\mathcal{\partial}_{em}^{0}\mathscr{V}_{g}^{0}+\mathcal{\partial}_{em}^{j}\mathscr{V}_{g}^{j}\right)\,\,= & \,\,0\,;\\
\left(\mathcal{\partial}_{em}^{0}\mathscr{V}_{em}^{0}+\mathcal{\partial}_{em}^{j}\mathscr{V}_{em}^{j}\right)\,= & \,\,0\,;\\
\left(\mathcal{\partial}_{g}^{0}\mathscr{V}_{g}^{0}+\mathcal{\partial}_{g}^{j}\mathscr{V}_{g}^{j}\right)\,\,= & \,\,0\,;\\
\left(\mathcal{\partial}_{g}^{0}\mathscr{V}_{em}^{0}+\mathcal{\partial}_{g}^{j}\mathscr{V}_{em}^{j}\right)\,= & \,\,0\,;
\end{split}
\label{eq:57}
\end{equation}
we get the following split octonion form of the field tensor associated
with the unified gravitational-electromagnetic (G-EM) interactions
as
\begin{align}
\mathscr{F}\,\,=\,\left(\mathscr{F}_{em}^{j}-\mathscr{F}_{g}^{j}\right)u_{j}^{*}+\left(\mathscr{F}_{em}^{j}+\mathscr{F}_{g}^{j}\right)u_{j}\,= & \left(\begin{array}{cc}
0 & -\left(\mathscr{F}_{em}^{j}+\mathscr{F}_{g}^{j}\right)e_{j}\\
\left(\mathscr{F}_{em}^{j}-\mathscr{F}_{g}^{j}\right)e_{j} & 0
\end{array}\right)\,.\label{eq:58}
\end{align}
which can further be reduced to
\begin{align}
\mathscr{F}\,\,= & \left(\begin{array}{cc}
0 & -\Psi_{(em-g)_{+}}^{j}e_{j}\\
\Psi_{(em-g)_{-}}^{j}e_{j} & 0
\end{array}\right)\,,\label{eq:59}
\end{align}
where $\Psi_{(em-g)_{+}}^{j}\rightarrow\left(\mathscr{F}_{em}^{j}+\mathscr{F}_{g}^{j}\right)$
and $\Psi_{(em-g)_{-}}^{j}\rightarrow\left(\mathscr{F}_{em}^{j}-\mathscr{F}_{g}^{j}\right)$
are generalized EM-G fields in octonionic space. Hence, we get the
split octonionic field equation in unified gravitational-electromagnetic
space on applying the differential operator (\ref{eq:50}) to equation
(\ref{eq:58}) as
\begin{align}
\mathfrak{D}\,\mathscr{F}\,\,= & \,\,\mathcal{J}\,\,,\label{eq:60}
\end{align}
where $\mathcal{J}$ is the octonionic form of unified gravitational-electromagnetic
field current defined as
\begin{align}
\mathcal{J}\,= & \left(\begin{array}{cc}
\left(\mathfrak{J}_{(em-g)}-\mathfrak{J}_{(em-em)}\right)e_{0} & -\left(\mathfrak{J}_{(em-em)/(g-g)}-\mathfrak{J}_{(em-g)/(g-em)}\right)e_{j}\\
\left(\mathfrak{J}_{(g-g)/(em-em)}+\mathfrak{J}_{(g-em)/(em-g)}\right)e_{j} & \left(\mathfrak{J}_{(g-em)}-\mathfrak{J}_{(g-g)}\right)e_{0}
\end{array}\right)\,.\label{eq:61}
\end{align}
Here $\left(\mathfrak{J}_{(em-g)}\,,\,\mathfrak{J}_{(em-em)}\right)$,
and $\left(\mathfrak{J}_{(g-em)}\,,\,\mathfrak{J}_{(g-g)}\right)$
are described as the octonionic current source density respectively
for (EM-G, EM-EM) and (G-EM, G-G) spaces, while other components $\left(\mathfrak{J}_{(em-em)/(g-g)}\,,\,\mathfrak{J}_{(em-g)/(g-em)}\right)$
and $\left(\mathfrak{J}_{(g-g)/(em-em)}\,,\,\mathfrak{J}_{(g-em)/(em-g)}\right)$
are the mixed current source densities of (EM-EM/G-G, EM-G/ G-EM)
and (G-G/EM-EM, G-EM/EM-G) octonionic spaces \cite{key-30}. Thus
from equation (\ref{eq:60}), we get
\begin{align}
\mathfrak{D}\,\mathscr{F}\,\,= & \left(\begin{array}{cc}
\left(\mathcal{\partial}_{em}^{j}\mathscr{F}_{g}^{j}-\mathcal{\partial}_{em}^{j}\mathscr{F}_{em}^{j}\right)e_{0} & -(\mathcal{\partial}_{em}^{0}\mathscr{F}_{em}^{k}-\epsilon_{ijk}\mathcal{\partial}_{g}^{i}\mathscr{F}_{g}^{j}-\\
\, & \mathcal{\partial}_{em}^{0}\mathscr{F}_{g}^{k}+\epsilon_{ijk}\mathcal{\partial}_{g}^{i}\mathscr{F}_{em}^{j})e_{k}\\
(-\mathcal{\partial}_{g}^{0}\mathscr{F}_{em}^{k}+\epsilon_{ijk}\mathcal{\partial}_{em}^{i}\mathscr{F}_{g}^{j}+ & \,\\
\mathcal{\partial}_{g}^{0}\mathscr{F}_{g}^{k}+\epsilon_{ijk}\mathcal{\partial}_{em}^{i}\mathscr{F}_{em}^{j})e_{k} & \left(\mathcal{\partial}_{g}^{j}\mathscr{F}_{em}^{j}-\mathcal{\partial}_{g}^{j}\mathscr{F}_{g}^{j}\right)e_{0}
\end{array}\right)\,.\label{eq:62}
\end{align}
where the components of split octonionic current source density $\mathcal{J}$
are described as
\begin{align}
\mathcal{\partial}_{em}^{j}\mathscr{F}_{g}^{j}e_{0}= & \,\,\mathfrak{J}_{(em-g)}^{j}\,;\nonumber \\
\mathcal{\partial}_{em}^{j}\mathscr{F}_{em}^{j}e_{0}= & \,\,\mathfrak{J}_{(em-em)}^{j}\,;\nonumber \\
\left(\mathcal{\partial}_{em}^{0}\mathscr{F}_{em}^{k}-\epsilon_{ijk}\mathcal{\partial}_{g}^{i}\mathscr{F}_{g}^{j}\right)e_{k}= & \,\,\mathfrak{J}_{(em-em)/(g-g)}^{k}\,;\nonumber \\
\left(\mathcal{\partial}_{em}^{0}\mathscr{F}_{g}^{k}+\epsilon_{ijk}\mathcal{\partial}_{g}^{i}\mathscr{F}_{em}^{j}\right)e_{k}= & \,\,\mathfrak{J}_{(em-g)/(g-em)}^{k}\,;\nonumber \\
\left(-\mathcal{\partial}_{g}^{0}\mathscr{F}_{em}^{k}+\epsilon_{ijk}\mathcal{\partial}_{em}^{i}\mathscr{F}_{g}^{j}\right)e_{k}= & \,\,\mathfrak{J}_{(g-em)/(em-g)}^{k}\,;\label{eq:63}\\
\left(\mathcal{\partial}_{g}^{0}\mathscr{F}_{g}^{k}+\epsilon_{ijk}\mathcal{\partial}_{em}^{i}\mathscr{F}_{em}^{j}\right)e_{k}= & \,\,\mathfrak{J}_{(g-g)/(em-em)}^{k}\,;\nonumber \\
\mathcal{\partial}_{g}^{j}\mathscr{F}_{em}^{j}e_{0}= & \,\,\mathfrak{J}_{(g-em)}^{j}\,;\nonumber \\
\mathcal{\partial}_{g}^{j}\mathscr{F}_{g}^{j}e_{0}= & \,\,\mathfrak{J}_{(g-g)}^{j}\,;\nonumber 
\end{align}
which are the various kinds of fields equations in presence of gravitational-gravitational
(G-G), electromagnetic-electromagnetic (EM-EM), electromagnetic-gravitational
(EM-G), gravitational-electromagnetic (G-EM) and their combined interactions.\\
Moreover, we write the split octonionic radius vector $\mathcal{R}$
as
\begin{align}
\mathcal{R}\,=\,\mathcal{R}_{em}^{0}u_{0}^{*}+\mathcal{R}_{g}^{0}u_{0}+\mathcal{R}_{em}^{j}u_{j}^{*}+\mathcal{R}_{g}^{j}u_{j}\,\,\simeq & \,\,\left(\begin{array}{cc}
\mathcal{R}_{em}^{0}e_{0} & -\mathcal{R}_{em}^{j}e_{j}\\
\mathcal{R}_{g}^{j}e_{j} & \mathcal{R}_{g}^{0}e_{0}
\end{array}\right),\label{eq:64}
\end{align}
which directly reproduces the velocity ($\mathfrak{V}$) of the particle
in unified gravitational-electromagnetic space. Furthermore, $\mathfrak{V}$
can be written as the following 2$\times$2 Zorn's matrix realization
of split octonion $(u_{0},u_{j},u_{0}^{*},u_{j}^{*})$ as
\begin{align}
\mathfrak{V}= & \,\,\frac{\partial\mathcal{R}}{\partial t}\,=\,\frac{\partial}{\partial t}\left\{ \mathcal{R}_{em}^{0}u_{0}^{*}+\mathcal{R}_{g}^{0}u_{0}+\mathcal{R}_{em}^{j}u_{j}^{*}+\mathcal{R}_{g}^{j}u_{j}\right\} \nonumber \\
= & \,\,\mathfrak{V}_{em}^{0}u_{0}^{*}+\mathfrak{V}_{g}^{0}u_{0}+\mathfrak{V}_{em}^{j}u_{j}^{*}+\mathfrak{V}_{g}^{j}u_{j},\nonumber \\
= & \,\,\left(\begin{array}{cc}
\mathfrak{V}_{em}^{0}e_{0} & -\mathfrak{V}_{em}^{j}e_{j}\\
\mathfrak{V}_{g}^{j}e_{j} & \mathfrak{V}_{g}^{0}e_{0}
\end{array}\right)\,.\label{eq:65}
\end{align}
Hence, the momentum of massive particle also be expressed in the following
octonionic form as
\begin{align}
\mathscr{P}\,\,=\,\,\mathscr{P}_{em}^{0}u_{0}^{*}+\mathscr{P}_{g}^{0}u_{0}+\mathscr{P}_{em}^{j}u_{j}^{*}+\mathscr{P}_{g}^{j}u_{j}\,= & \left(\begin{array}{cc}
\mathscr{P}_{em}^{0}e_{0} & -\mathscr{P}_{em}^{j}e_{j}\\
\mathscr{P}_{g}^{j}e_{j} & \mathscr{P}_{g}^{0}e_{0}
\end{array}\right)\,,\label{eq:66}
\end{align}
whereas the charge and mass \cite{key-30} of the particle in octonion
formulation may also be analysed as,
\begin{align}
\mathfrak{J}_{(em-em)/(g-g)}\,= & \,\, Q_{(em-em)/(g-g)}\,\mathfrak{V}_{(em-em)/(g-g)}\,,\nonumber \\
\mathfrak{J}_{(em-g)/(g-em)}\,= & \,\, Q_{(em-g)/(g-em)}\,\mathfrak{V}_{(em-g)/(g-em)}\,,\label{eq:67}\\
\mathfrak{J}_{g-g}=\,\, Q_{\left(g-g\right)}\mathfrak{V}_{(g-g)} & ,\,\,\,\mathfrak{J}_{(em-em)}=\,\, Q_{\left(em-em\right)}\mathfrak{V}_{(em-em)}\,,\nonumber 
\end{align}
where $Q_{\left(g-g\right)},\, Q_{\left(em-g\right)},\, Q_{\left(g-em\right)}$
and $Q_{(em-g)/(g-em)}$, $Q_{em-em)/(g-g)}$ respectively denote
the \textit{mass} of the gravitational- gravitational (G-G), electromagnetic-
gravitational (EM-G), gravitational- electromagnetic (G-EM) interactions
etc., while $Q_{\left(em-em\right)}$ represent the \textit{charge}
of the electromagnetic- electromagnetic (EM-EM) interaction. Thus
$Q_{\left(g-g\right)},$ $Q_{\left(em-g\right)},\, Q_{\left(g-em\right)}$,
their other combinations and $Q_{\left(em-em\right)}$ respectively
describe the \textit{generalized mass} and \textit{generalized charge}
\cite{key-30,key-35,key-36,key-37}of dyons in the consistent unified
picture for split octonions.

\section{Octonion Dark Matter (ODM)}

Dark matter \cite{key-38,key-39} is a type of matter hypothesized
to account for a large part of the total mass in the universe. Only
about 4.6\% of the mass-energy of the universe is ordinary matter,
about 23\% is thought to be composed of dark matter. The remaining
72\% is thought to consist of dark energy, an even stranger component,
distributed almost uniformly in space and with energy density non-evolving
or slowly evolving with time. On the other hand, the nonbaryonic form
of octonionic dark matter \cite{key-38,key-39} is evident through
its gravitational effect only. 

However, form experiment point of views the summary of the current
measurements of the matter density $\Omega_{matter}$, the dark matter
density $\Omega_{DM}$ and the energy density $\Omega_{energy}$ is
already given by Hagiwara \cite{key-41}. Each are in units of the
critical density $\lyxmathsym{\textgreek{r}}_{critical}\simeq3h^{2}/(8\lyxmathsym{\textgreek{p}}\mathbb{G})$,
where $\mathbb{G}$ is the Newton\textquoteright{}s gravitational
constant, and $h$ is the present value of the Hubble constant. The
experimental results \cite{key-42,key-43} require the best current
values of the matter and energy densities respectively given as $\Omega_{matter}\simeq$
0.27 and $\Omega_{energy}\simeq$ 0.73. The measurements describe
the baryon density $\Omega_{baryon}$ to a value less than $\sim$
0.03. The difference $\Omega_{matter}-\Omega_{baryon}\simeq$ 0.24
must be in the form of non-baryonic dark matter. Some experimental
results for energy, matter and dark matter densities \cite{key-44,key-45}
are shown in Table-2.

\begin{table}
\centering{}%
\begin{tabular}{c|c}
\hline 
$0.49\leq\Omega_{energy}\leq0.74$ & $0.49\leq\Omega_{energy}\leq0.76$\tabularnewline[0.2cm]
$0.20\leq\Omega_{matter}\leq0.50$ & \tabularnewline[0.2cm]
$0.11\leq h^{2}\Omega_{DM}\leq0.17$ & $0.09\leq h^{2}\Omega_{DM}\leq0.17$\tabularnewline[0.2cm]
$0.00\leq h^{2}\Omega_{Hot-dark-matter}\leq0.12$ & \tabularnewline[0.2cm]
$0.10\leq h^{2}\Omega_{Cold-dark-matter}\leq0.32$ & \tabularnewline[0.2cm]
$0.02\leq h^{2}\Omega_{baryon}\leq0.03$ & $0.01\leq h^{2}\Omega_{baryon}\leq0.03$\tabularnewline[0.2cm]
\hline 
\end{tabular}\caption{The experimental results for Matter and Dark Matter densities with
the compression of first release data and newest data. }
\end{table}
The ranges of density parameters within the standard cosmological
model derived from the first release data with corresponding values
newest data and $h\simeq0.74\pm0.08$ is the Hubble parameter \cite{key-44,key-45}.
Thus, the matter density: 
\begin{eqnarray}
\Omega_{matter}\,\, & \simeq & \,\,\Omega_{baryon}+\Omega_{Hot-dark-matter}+\Omega_{Cold-dark-matter}\nonumber \\
\, & \simeq & \,\,\Omega_{baryon}+\Omega_{DM}\,,\label{eq:68}
\end{eqnarray}
includes both baryonic matter and dark matter. Moreover the dark matter
can be classed either as \textit{hot} or \textit{cold} depending on
whether it was relativistic or not in the early universe. So, we can
analyse the hot and cold dark matter in terms of split octonions.
We assume that the total octonionic dark matter density (ODM) is described
as 
\begin{align}
\Omega_{\mathcal{O}-DM}\,\, & \simeq\,\,\Omega_{\mathcal{O}-HDM}+\Omega_{\mathcal{O}-CDM}\,\,,\label{eq:69}
\end{align}
where $\Omega_{\mathcal{O}-HDM}$ describes the octonionic hot dark
matter density (OHDM) while $\Omega_{\mathcal{O}-CDM}$ describes
the octonionic cold dark matter density (OCDM).

\section{Octonion Hot and Cold Dark matter}

Let us assumed that the octonionic hot dark matter (OHDM) describes
the particles that have zero or near-zero masses. Thus we assume that
the octonionic hot dark matter (OHDM) is associated with the gravitational-gravitational
(G-G), electromagnetic-electromagnetic (EM-EM) and the combination
of (G-G) and (EM-EM) fields \cite{key-30,key-35,key-36,key-37} interactions.
So, the hot dark matter may then be described in terms of following
2$\times$2 Zorn vector matrix realization of split octonions, i.e.
\begin{align}
\mathscr{F}_{\mathcal{O}-HDM}\,\,= & \left(\begin{array}{cc}
0 & (\mathcal{\partial}_{em}^{k}\mathscr{V}_{em}^{0}+\mathcal{\partial}_{em}^{0}\mathscr{V}_{em}^{k}-\epsilon_{ijk}\mathcal{\partial}_{g}^{i}\mathscr{V}_{g}^{j})e_{k}\\
(-\mathcal{\partial}_{g}^{k}\mathscr{V}_{g}^{0}-\mathcal{\partial}_{g}^{0}\mathscr{V}_{g}^{k}+\epsilon_{ijk}\mathcal{\partial}_{em}^{i}\mathscr{V}_{em}^{j})e_{k} & 0
\end{array}\right)\,,\label{eq:70}
\end{align}
where $\mathscr{F}_{\mathcal{O}-HDM}$ is an octonionic field strength
represents the hot dark matter. So, the components of equation (\ref{eq:70})
are given as
\begin{equation}
\begin{cases}
\mathscr{F}_{(em-em)/(g-g)}^{k} & =\,\,\left(\mathcal{\partial}_{em}^{k}\mathscr{V}_{em}^{0}+\mathcal{\partial}_{em}^{0}\mathscr{V}_{em}^{k}-\epsilon_{ijk}\mathcal{\partial}_{g}^{i}\mathscr{V}_{g}^{j}\right)e_{k}\,;\\
\mathscr{F}_{(g-g)/(em-em)}^{k} & =\,\,\left(-\mathcal{\partial}_{g}^{k}\mathscr{V}_{g}^{0}-\mathcal{\partial}_{g}^{0}\mathscr{V}_{g}^{k}+\epsilon_{ijk}\mathcal{\partial}_{em}^{i}\mathscr{V}_{em}^{j}\right)e_{k}\,.
\end{cases}\label{eq:71}
\end{equation}
Here the octonion hot dark matter (OHDM) represents the quantas of
EM-EM interactions (namely photons) and G-G interactions (i.e. gravitons).
Similarly, the required current source density for octonionic hot
dark matter (OHDM) is expressed by the following split octonion form
as,
\begin{align}
\mathcal{J}_{\mathcal{O}-HDM}\,\,= & \begin{pmatrix}0 & -(\mathcal{\partial}_{em}^{0}\mathscr{F}_{em}^{k}-\epsilon_{ijk}\mathcal{\partial}_{g}^{i}\mathscr{F}_{g}^{j})e_{k}\\
(\mathcal{\partial}_{g}^{0}\mathscr{F}_{g}^{k}+\epsilon_{ijk}\mathcal{\partial}_{em}^{i}\mathscr{F}_{em}^{j})e_{k} & 0
\end{pmatrix}\,,\label{eq:72}
\end{align}
which represent the non-baryonic octonion hot dark matter current
source density $\mathcal{J}_{\mathcal{O}-HDM}$ associated with massless
photons, gravitons, etc. Thus the current source equations for octonionic
hot dark matter (OHDM) are given below:
\begin{equation}
\begin{cases}
\mathfrak{J}_{(em-em)/(g-g)}^{k} & =\,\,\left(\mathcal{\partial}_{em}^{0}\mathscr{F}_{em}^{k}-\epsilon_{ijk}\mathcal{\partial}_{g}^{i}\mathscr{F}_{g}^{j}\right)e_{k}\,;\\
\mathfrak{J}_{(g-g)/(em-em)}^{k} & =\,\,\left(\mathcal{\partial}_{g}^{0}\mathscr{F}_{g}^{k}+\epsilon_{ijk}\mathcal{\partial}_{em}^{i}\mathscr{F}_{em}^{j}\right)e_{k}\,.
\end{cases}\label{eq:73}
\end{equation}
Rather, the octonion cold dark matter (OCDM) can be described as the
composition of the massive objects moving at sub-relativistic velocities.
So, the difference between the octonion cold dark matter (OCDM) and
the octonions hot dark matter (OHDM) is significant in the formulation
of structure, because the velocities of octonions hot dark matter
cause it to wipe out structure on small scales. Thus, the octonions
cold dark matter is associated with the electromagnetic-gravitational
(EM-G), gravitational-electromagnetic (G-EM) and their mixed interactions.
Thus, we analyse the octonion cold dark matter for massive particles,
responsible for EM-G and G-EM interactions, i.e. 
\begin{align}
\mathscr{F}_{\mathcal{O}-CDM}\,\,= & \left(\begin{array}{cc}
0 & (\mathcal{\partial}_{em}^{k}\mathscr{V}_{g}^{0}+\mathcal{\partial}_{em}^{0}\mathscr{V}_{g}^{k}+\epsilon_{ijk}\mathcal{\partial}_{g}^{i}\mathscr{V}_{em}^{j})e_{k}\\
(\mathcal{\partial}_{g}^{k}\mathscr{V}_{em}^{0}+\mathcal{\partial}_{g}^{0}\mathscr{V}_{em}^{k}+\epsilon_{ijk}\mathcal{\partial}_{em}^{i}\mathscr{V}_{g}^{j})e_{k} & 0
\end{array}\right)\,\,,\label{eq:74}
\end{align}
where $\mathscr{F}_{\mathcal{O}-CDM}$ represents the octonionic field
tensor associated with the cold dark matter. Equation (\ref{eq:74})
reduced to following components 
\begin{equation}
\begin{cases}
\mathscr{F}_{(em-g)/(g-em)}^{k} & =\,\,\left(\mathcal{\partial}_{em}^{k}\mathscr{V}_{g}^{0}+\mathcal{\partial}_{em}^{0}\mathscr{V}_{g}^{k}+\epsilon_{ijk}\mathcal{\partial}_{g}^{i}\mathscr{V}_{em}^{j}\right)e_{k}\,,\\
\mathscr{F}_{(g-em)/(em-g)}^{k} & =\,\,\left(\mathcal{\partial}_{g}^{k}\mathscr{V}_{em}^{0}+\mathcal{\partial}_{g}^{0}\mathscr{V}_{em}^{k}+\epsilon_{ijk}\mathcal{\partial}_{em}^{i}\mathscr{V}_{g}^{j}\right)e_{k}\,.
\end{cases}\label{eq:75}
\end{equation}
So, the mediators for octonionic cold dark matter (OCDM), which are
non-baryonic particles like $W^{\pm},\, Z^{0},$ etc. Therefore, the
current source density for octonionic cold dark matter (OCDM) can
expressed as 
\begin{align}
\mathcal{J}_{\mathcal{O}-CDM}\,\,= & \begin{pmatrix}\left(\mathcal{\partial}_{em}^{j}\mathscr{F}_{g}^{j}-\mathcal{\partial}_{em}^{j}\mathscr{F}_{em}^{j}\right)e_{0} & -\mathcal{\partial}_{em}^{0}\mathscr{F}_{g}^{k}+\epsilon_{ijk}\mathcal{\partial}_{g}^{i}\mathscr{F}_{em}^{j})e_{k}\\
(-\mathcal{\partial}_{g}^{0}\mathscr{F}_{em}^{k}+\epsilon_{ijk}\mathcal{\partial}_{em}^{i}\mathscr{F}_{g}^{j})e_{k} & \left(\mathcal{\partial}_{g}^{j}\mathscr{F}_{em}^{j}-\mathcal{\partial}_{g}^{j}\mathscr{F}_{g}^{j}\right)e_{0}
\end{pmatrix}\,,\label{eq:76}
\end{align}
where $\mathcal{J}_{\mathcal{O}-CDM}$ defines the octonionic cold
dark matter current source density, which associated EM-G and EM-G
fields spaces. Thus the current source equations for octonionic cold
dark matter are:
\begin{equation}
\begin{cases}
\mathfrak{J}_{(em-g)/(em-em)}^{j} & =\,\,\left(\mathcal{\partial}_{em}^{j}\mathscr{F}_{g}^{j}-\mathcal{\partial}_{em}^{j}\mathscr{F}_{em}^{j}\right)\,;\\
\mathfrak{J}_{(g-em)/(g-g)}^{j} & =\,\,\left(\mathcal{\partial}_{g}^{j}\mathscr{F}_{em}^{j}-\mathcal{\partial}_{g}^{j}\mathscr{F}_{g}^{j}\right)\,;\\
\mathfrak{J}_{(em-g)/(g-em)}^{k} & =\,\,\left(\mathcal{\partial}_{em}^{0}\mathscr{F}_{g}^{k}+\epsilon_{ijk}\mathcal{\partial}_{g}^{i}\mathscr{F}_{em}^{j}\right)e_{k}\,;\\
\mathfrak{J}_{(g-em)/(em-g)}^{k} & =\,\,\left(-\mathcal{\partial}_{g}^{0}\mathscr{F}_{em}^{k}+\epsilon_{ijk}\mathcal{\partial}_{em}^{i}\mathscr{F}_{g}^{j}\right)e_{k}\,.
\end{cases}\label{eq:77}
\end{equation}
So, the quantum equations for octonionic hot and cold dark matter
may easily be expressed in the terms of simpler and compact notation
of octonions representations. From the foregoing analysis we find
the matter density of octonionic cold dark matter (OCDM) i.e. $\left(0.10\leq h^{2}\Omega_{CDM}\leq0.32\right)$
is grater than that of octonionic hot dark matter (OHDM) as $\left(0.00\leq h^{2}\Omega_{HDM}\leq0.12\right)$.
As such, there exists more cold dark matter than that of hot dark
matter in nature supported by octonionic field equations.


\begin{thebibliography}{10}
\bibitem[1]{key-1} L. E. Dickson, \textbf{\textquotedblleft{}On Quaternions
and Their Generalization and the History of the Eight Square Theorem\textquotedblright{}},
Ann. Math., \textbf{\uline{20}} (1919), 155.

\bibitem[2]{key-2} W. R. Hamilton, \textbf{\textquotedblleft{}Elements
of Quaternions\textquotedblright{}}, Chelsea Publications Co., New
York, (1969). 

\bibitem[3]{key-3} P. G. Tait, \textbf{\textquotedbl{}An elementary
Treatise on Quaternions\textquotedbl{},} Oxford Univ. Press (1875).

\bibitem[4]{key-4} A. Cayley, \textbf{\textquotedblleft{}An Jacobi\textquoteright{}s
elliptic functions, in reply to the Rev. B. Bornwin; and on quaternion\textquotedblright{}},
Phil. Mag., \textbf{\uline{26}} (1845), 208.

\bibitem[5]{key-5} R. P. Graves, \textbf{\textquotedblleft{}Life
of Sir William Rowan Hamilton\textquotedblright{},} 3 volumes, Arno
Press, New York, (1975).

\bibitem[6]{key-6} J. C. Baez, \textbf{\textquotedblleft{}The Octonions\textquotedblright{}},
Bull. Amer. Math. Soc., \textbf{\uline{39}} (20R. Bond and A. Szalay,
Astrophys. J. 274 (1983) 443.01), 145.

\bibitem[7]{key-7} K. Morita, \textbf{\textquotedblleft{}Quaternionic
Variational Formalism for Poincaré Gauge Theory and Supergravity\textquotedblright{}},
Prog. Theor. Phys., \textbf{\uline{73}} (1985), 999. 

\bibitem[8]{key-8} K. Morita, \textbf{\textquotedblleft{}Gauge Theories
over Quaternions and Weinberg-Salam Theory\textquotedblright{},} Prog.
Theor. Phys., \textbf{\uline{65}} (1981), 2071. 

\bibitem[9]{key-9} K. Morita, \textbf{\textquotedblleft{}Quaternionic
Weinberg-Salam Theory\textquotedblright{},} Prog. Theor. Phys., \textbf{\uline{67}}
(1982), 1860. 

\bibitem[10]{key-10} S. Catto, \textbf{\textquotedblleft{} Exceptional
Projective Geometries and Internal Symmetries\textquotedblright{}},
eprint: arXiv: hep-th/0302079 (2003).

\bibitem[11]{key-11} S. Catto,\textbf{ \textquotedblleft{}Symmetries
in Science:from the rotation group to quantum algebras\textquotedblright{}},
Ed. B. Gruber, Plenum Press, \textbf{\uline{6}} (1993), 129.

\bibitem[12]{key-12} R. Foot and G. C. Joshi, \textbf{\textquotedblleft{}Space-time
symmetries of superstring and Jordan Algebras\textquotedblright{}},
Int. J. Theor. Phys.,\textbf{\uline{ 28}} (1989), 1449.

\bibitem[13]{key-13} R. Foot and G. C. Joshi, \textbf{\textquotedblleft{}String
theories and the Jordan algebras\textquotedblright{}}, Phys. Lett.,
\textbf{\uline{B199 }}(1987), 203.

\bibitem[14]{key-14} T. Kugo and P. Townsend, \textbf{\textquotedblleft{}Supersymmetry
and the Division Algebras\textquotedblright{}}, Nucl Phys., \textbf{\uline{B221}}
(1983), 357.

\bibitem[15]{key-15} C. A. Manogue and A. Sudbery, \textbf{\textquotedblleft{}General
solutions of covariant superstring equations of motion\textquotedblright{}},
Phys. Rev., \textbf{\uline{D40}} (1989), 4073.

\bibitem[16]{key-16} J. Schray, \textbf{\textquotedblleft{}Octonions
and Supersymmetry\textquotedblright{},} Ph. D. thesis, Department
of Physics, Oregon State University, (1994) (un published).

\bibitem[17]{key-17} K. S. Abdel-Khalek Mostafa, \textbf{\textquotedblleft{}Ring
Division Algebra, Self Duality and Supersymmetry\textquotedblright{}},
eprint: arXiv: hep-th/0002155 (2000).

\bibitem[18]{key-18} M. Günaydin and F. Gürsey, \textbf{\textquotedblleft{}Quark
structure and octonions\textquotedblright{}}, J. Math. Phys., \textbf{\uline{14}}
(1973), 1651.

\bibitem[19]{key-19} K. Imaeda, \textbf{``Quaternionic formulation
of tachyons, superluminal transformations and a complex space-time''},
Lett. Nuovo Cimento, \textbf{\uline{50}} (1979), 271.

\bibitem[20]{key-20} R. Penny, \textbf{``Octonions and the Dirac
equation''}, Amer. J. Phys., \textbf{\uline{36}} (1968), 871.

\bibitem[21]{key-21} A. Gamba, \textbf{``Peculiarities of the Eight
Dimensional Space''}, J. Math. Phys., \textbf{\uline{8}} (1967),
775.

\bibitem[22]{key-22}\textcolor{black}{{} B. C. Chanyal, P. S. Bisht
and O. P. S. Negi,}\textit{ }\textbf{``Generalized Octonion Electrodynamics'',}
Int. J. Theor. Phys., \textbf{\uline{49}} (2010), 1333.

\bibitem[23]{key-23} \textcolor{black}{B. C. Chanyal, P. S. Bisht
and O. P. S. Negi,} \textbf{``Generalized Split-Octonion Electrodynamics'',}\textit{
}Int. J. Theor. Phys., \textbf{\uline{50}} (2011), 1919.

\bibitem[24]{key-24}\textcolor{black}{{} B. C. Chanyal, P. S. Bisht,
Tianjun Li and O. P. S. Negi,} \textbf{``Octonion Quantum Chromodynamics'',}
Int. J. Theor. Phys., \textbf{\uline{51}} (2012), 3410.

\bibitem[25]{key-25} \textcolor{black}{B. C. Chanyal, P. S. Bisht
and O. P. S. Negi,} \textbf{\textquotedblleft{}Octonionic non-Abelian
Gauge Theory\textquotedblright{},} Int. J. Theor. Phys.,\textcolor{black}{{}
}\textbf{\textcolor{black}{\uline{52}}}\textcolor{black}{{} (2013),
3522.}

\bibitem[26]{key-26} \textcolor{black}{B. C. Chanyal, P. S. Bisht
and O. P. S. Negi, }\textbf{\textcolor{black}{``Octonion and conservation
laws for dyons'',}}\textcolor{black}{{} Int. J. Mod. Phys. A, }\textbf{\textcolor{black}{\uline{28}}}\textcolor{black}{{}
(2013), 1350125.}

\bibitem[27]{key-27}\textcolor{black}{{} B. C. Chanyal, P. S. Bisht
and O. P. S. Negi,}\textbf{\textcolor{black}{{} ``Octonion Electrodynamics
in Isotropic and Chiral Medium'', }}\textcolor{black}{Int. J. Mod.
Phys. A, }\textbf{\textcolor{black}{\uline{29}}}\textcolor{black}{{}
(2014), 1450008.}

\bibitem[28]{key-28} \textcolor{black}{B. C. Chanyal, }\textbf{``Octonion
massive electrodynamics'',}\textit{\textcolor{black}{{} }}\textcolor{black}{Gen.
Relativ. Gravit., }\textbf{\textcolor{black}{\uline{46}}}\textcolor{black}{{}
(2014), 16461.}

\bibitem[29]{key-29}\textcolor{black}{{} B. C. Chanyal, P. S. Bisht
and O. P. S. Negi,}\textbf{\textcolor{black}{{} ``Octonion Representation
of the Superstring Theory''}}\textcolor{black}{{} Int. J. Eng. Res.
\& Tech., }\textbf{\textcolor{black}{\uline{2}}}\textcolor{black}{{}
(2013), 1459.}

\bibitem[30]{key-30}\textcolor{black}{{} B. C. Chanyal, P. S. Bisht
and O. P. S. Negi, }\textbf{\textcolor{black}{``Octonion model of
dark matter}}\textcolor{black}{'', J. Theor. Phys., }\textbf{\textcolor{black}{\uline{2}}}\textcolor{black}{{}
(2013), 23.}

\bibitem[31]{key-31}\textcolor{black}{{} B. C. Chanyal, ``}\textbf{\textcolor{black}{Octonion
symmetric Dirac-Maxwell equations}}\textcolor{black}{'', Turk. J.
Phys., }\textbf{\textcolor{black}{\uline{38}}}\textcolor{black}{{}
(2014), 174.}

\bibitem[32]{key-32} M. Günaydin and F. Gursey, \textbf{``Quark
structure and octonions'',} J. Math. Phys., \textbf{\uline{14}}
(1973), 1651.

\bibitem[33]{key-33} M. Günaydin and F. Gursey, \textbf{``Quark
statistics and octonions'',} Phys. Rev. D, \textbf{\uline{9}}
(1974), 3387.

\bibitem[34]{key-34} S. Marques and C. G. Oliveira, ``\textbf{Geometrical
properties of an internal local octonionic space in curved space time}'',
Phys. Rev. D, \textbf{\uline{36}} (1987), 1716.

\bibitem[35]{key-35} Z. Weng, ``\textbf{Octonionic Quantum Interplays
of Dark Matter and Ordinary Matter}'', arXiv: physics.gen-ph/0702019v4,
(2008).

\bibitem[36]{key-36} Z. Weng, ``\textbf{Octonionic electromagnetic
and gravitational interactions and dark matter}'', arXiv: physics.class-ph/0612102v8,
(2009).

\bibitem[37]{key-37} \textcolor{black}{B. C. Chanyal, ``}\textbf{\textcolor{black}{Role
of Octonion in gravity and dark matter}}\textcolor{black}{'', Cliff.
Analy. Cliff. Algeb. \& Appl., }\textbf{\textcolor{black}{\uline{3}}}\textcolor{black}{{}
(2014), 121.}

\bibitem[38]{key-38} G. Bertone, D. Hooper and J. Silk, ``\textbf{Particle
darkmatter: evidence, candidates and constraints}'', Physics Reports,
Vol. \textbf{\uline{405}} (2005), 279.

\bibitem[39]{key-39} M. Dine, ``\textbf{Origin of the matter-antimatter
asymmetry}'', Rev. Mod. Phys., \textbf{\uline{76}} (2003), 1.

\bibitem[40]{key-40} R. Bond and A. Szalay, ``\textbf{The collisionless
damping} \textbf{of density fluctuations in an expanding universe}''
Astrophys. J. \textbf{\uline{274}} (1983), 443.

\bibitem[41]{key-41} K Hagiwara, et al ``\textbf{Review of Particle
Physics. Particle Data Group}'' Phys. Rev. \textbf{\uline{D66}}
(2002), 010001.

\bibitem[42]{key-42} S P Ahlen, et al ``\textbf{Limits on Cold}
\textbf{Dark Matter Candidates from an Ultralow Background Germanium
Spectrometer}'', Phys. Lett.\textit{,} \textbf{\uline{B195}} (1987),
603.

\bibitem[43]{key-43} D S Akerib, et al ``\textbf{New results from
the cryogenic dark matter search experiment}'', Phys. Rev\textit{.,}
\textbf{\uline{D68}} (2003), 082002.

\bibitem[44]{key-44} M Tegmark, et al \textbf{``Sources and Detection
of Dark Matter and Dark Energy in the Universe''}, Springer, Berlin
(2000), (hep-ph/0008145).

\bibitem[45]{key-45} X Wang, et al ``\textbf{Is cosmology consistent?}'',
Phys. Rev., \textbf{\uline{D65}} (2002), 123001.\end{thebibliography}
\end{document}